\documentclass[11pt,epsf]{article}
\usepackage{amsmath}
\usepackage{amsfonts}
\usepackage{amssymb}
\usepackage{graphicx}
\newtheorem{theorem}{Theorem}

\newcommand {\dfn} {\stackrel{\Delta} {=}}

\newcommand {\bx} {\mbox{\boldmath $x$}}
\newcommand {\by} {\mbox{\boldmath $y$}}

\newcommand {\bE} {\mbox{\boldmath $E$}}

\newcommand {\bU} {\mbox{\boldmath $U$}}

\newcommand {\bX} {\mbox{\boldmath $X$}}
\newcommand {\bY} {\mbox{\boldmath $Y$}}

\newcommand{\calA}{{\cal A}}
\newcommand{\calB}{{\cal B}}

\newcommand{\calE}{{\cal E}}

\newcommand{\calU}{{\cal U}}

\newcommand{\calX}{{\cal X}}
\newcommand{\calY}{{\cal Y}}

\begin{document}
\thispagestyle{empty}
\title{Exponential Error Bounds on Parameter Modulation--Estimation
for Discrete Memoryless Channels
}
\author{Neri Merhav
}
\date{}
\maketitle

\begin{center}
Department of Electrical Engineering \\
Technion - Israel Institute of Technology \\
Technion City, Haifa 32000, ISRAEL \\
E--mail: {\tt merhav@ee.technion.ac.il}\\
\end{center}
\vspace{1.5\baselineskip}
\setlength{\baselineskip}{1.5\baselineskip}

\begin{abstract}
We consider the problem of modulation and estimation of a random parameter $U$ to be
conveyed across a discrete memoryless channel. Upper and lower bounds are
derived for the
best achievable exponential decay rate of a general moment of the estimation
error, $\bE|\hat{U}-U|^\rho$, $\rho\ge 0$, when both the modulator and the
estimator are subjected to optimization.
These exponential error bounds turn out to be intimately related to error
exponents of channel coding and to channel capacity. 
While in general, there is some gap between the upper
and the lower bound, they
asymptotically coincide both for very small and for very large values
of the moment power $\rho$. This means that our achievability scheme,
which is based on simple quantization of $U$ followed by channel coding, is nearly
optimum in both limits. Some additional properties of the bounds are discussed
and demonstrated, and finally, an extension to the case of a multidimensional parameter
vector is outlined, with the principal conclusion that our upper and lower
bound asymptotically coincide also for a high dimensionality.

\vspace{0.4cm}

\end{abstract}

\newpage
\section{Introduction}

Consider the problem of conveying the value of a parameter $u$ across a
given discrete memoryless channel
\begin{equation}
p(\by|\bx)=\prod_{t=1}^n p(y_t|x_t),
\end{equation}
where $\bx=(x_1,\ldots,x_n)$ and
$\by=(y_1,\ldots,y_n)$ are the channel input and output vectors, respectively. 
Our main interest, in this work, is in the following questions: How well can one estimate $u$
based on $\by$ when one is allowed to optimize, not only the estimator, but
also the modulator, that is, the function $\bx(u)=(x_1(u),\ldots,x_n(u))$ that maps
$u$ into a channel
input vector? How fast does the estimation error decay as a function of $n$
when the best modulator and estimator are used?

In principle, this problem, which is the discrete--time analogue of the 
classical problem of ``waveform communication'' (in the terminology of
\cite[Chap.\ 8]{WJ65}), 
can be viewed both from the information--theoretic and the
estimation--theoretic perspectives. Classical results in neither of these
disciplines, however, seem to suggest satisfactory answers. 

From the information--theoretic point of
view, if the parameter is random, call it $U$,
this is actually a problem of joint source--channel coding, where the source emits
a single variable $U$ (or a fixed number of them when $U$ is a vector), 
whereas the channel is allowed to be used many times
($n$ is large). The separation theorem of classical information theory asserts
that asymptotic optimality of separate source-- and channel
coding is guaranteed in the limit of long blocks. However, it refers to a regime
of long blocks both in source coding and channel coding, whereas here the
source block length is 1, and so, there is no hope to compress the source 
with performance that comes close to the rate--distortion function.

In the realm of estimation theory, on the other hand, there is a rich
literature on Bayesian and non--Bayesian bounds, mostly concerning
the mean square error (MSE) in estimating parameters from signals
corrupted by an additive white Gaussian noise (AWGN) channel, as well as other
channels (see, e.g., \cite{TB07} and the introductions of \cite{BSET97},
\cite{BE09}, and \cite{Weiss85}
for overviews on these bounds). Most of these bounds lend themselves
to calculation for
a {\it given} modulator $\bx(u)$ and
therefore they may give insights concerning optimum
estimation for this specific modulator.
They may not, however, be easy to use for the derivation of
{\it universal} lower bounds, namely, lower bounds that
depend neither on the modulator nor on the estimator, which are relevant
when both optimum modulators and optimum
estimators are sought. 
Two exceptions to this rule
(although usually, not presented as such)
are families of bounds
that stem from generalized data processing theorems (DPT's) \cite{LZ07}, \cite{p146},
\cite{TZ11}, \cite{ZZ75}, \cite{ZZ73}, henceforth referred to as ``DPT
bounds'', and bounds
based on hypothesis testing
and channel coding considerations \cite{BSET97}, \cite{CZZ75}, \cite{ZZ69},
henceforth called ``channel--coding bounds.''

In this paper, we use both the channel--coding techniques and DPT techniques in
order to derive lower bounds on general moments of the estimation error,
$\bE|\hat{U}-U|^\rho$, where $U$ is a random parameter, $\hat{U}$ is its
estimate, and the power $\rho$ is an arbitrary positive real (not necessarily
an integer). It turns out that when $\bx(u)$ is subjected to optimization,
$\bE|\hat{U}-U|^\rho$ can decay exponentially rapidly as a function of $n$, 
and so, our focus is on the best achievable exponential rate of decay as a
function of $\rho$, which we shall denote by $\calE(\rho)$, that is,
\begin{equation}
\inf \bE|\hat{U}-U|^\rho \approx e^{-n\calE(\rho)},
\end{equation}
where the infimum is over all modulators and estimators.\footnote{This is
still an informal and non--rigorous description. More precise definitions will be given
in the sequel.}
Interestingly, both the upper and lower bounds on $\calE(\rho)$ are intimately
related to well--known exponential error bounds associated with channel
coding, such as Gallager's random coding exponent (for small values of $\rho$)
and the expurgated exponent
function (for large values of $\rho$). 
In other words, we establish an estimation--theoretic meaning to these
error exponent functions. In particular, under certain conditions, our 
channel--coding upper bound on $\calE(\rho)$ 
(corresponding to a lower bound on $\bE|\hat{U}-U|^\rho$) can be presented as
\begin{equation}
\overline{E}(\rho)=\left\{\begin{array}{ll}
E_0(\rho) & \rho < \rho_0\\
E_{ex}(0) & \rho \ge \rho_0
\end{array}\right.
\end{equation}
where $E_0(\rho)=\max_q E_0(\rho,q)$, $E_0(\rho,q)$ being Gallager's function,
$E_{ex}(0)$ is the expurgated exponent at zero rate, and $\rho_0$ is value of
$\rho$ for which $E_0(\rho)=E_{ex}(0)$ (so that $\overline{E}(\rho)$ is
continuous). In addition,
we derive a DPT bound and discuss its advantages
and disadvantages compared to the above bound.

We also suggest a lower bound, $\underline{E}(\rho)$, on $\calE(\rho)$ (associated with upper bounds on
$\inf \bE|\hat{U}-U|^\rho $), which is achieved by a simple, separation--based
modulation and estimation scheme. While there is a certain gap between
$\overline{E}(\rho)$ and $\underline{E}(\rho)$ for every finite $\rho$, it
turns out that this gap disappears (in the sense that the ratio
$\underline{E}(\rho)/\overline{E}(\rho)$ tends to
unity) both for large $\rho$ and for small $\rho$, and so, we have
exact asymptotics of $\calE(\rho)$ in these two extremes: For large $\rho$,
$\calE(\rho)$ tends to $E_{ex}(0)$ and for small $\rho$, $\calE(\rho)\sim \rho
C$, where $C$ is the channel capacity. Our simple achievability scheme is then
nearly optimum at both extremes, which means that a separation theorem 
essentially holds for very small and for very large values of $\rho$, in spite of
the earlier discussion (see also \cite[Section III.D]{p152}). The
results are demonstrated for the example of a ``very noisy channel,''
\cite[Example 3, pp.\ 147--149]{Gallager68}, \cite[pp.\ 155--158]{VO79},
which is convenient to analyze, as it admits closed--form expressions.

Finally, we suggest an extension of our results to the case of a
multidimensional parameter vector $\bU=(U_1,\ldots,U_d)$. It turns out that
the effect of the dimension $d$ is in reducing the effective value of $\rho$ by a factor of $d$. In
other words, $\overline{E}(\rho)$ is replaced by 
$\overline{E}(\rho/d)$ and the extension of the achievability result is
straightforward. This means that for fixed $\rho$, the limit of large $d$
(where the effective value $\rho/d$ is very small) also admits exact asymptotics, where
$\calE(\rho)\sim \rho C/d$.

The outline of the paper is as follows. 
In Section 2, we define the problem formally and we establish notation
conventions. In Section 3, we derive our main
upper and lower bounds based on channel coding considerations.
In Section 4, we derive our DPT bound and discuss it. Section 5 is
devoted to the example of the very noisy channel, and finally, in Section 6
the multidimensional case is considered.

\section{Notation Conventions and Problem Formulation}

Throughout this paper, scalar random
variables (RV's) will be denoted by capital
letters, their sample values will be denoted by
the respective lower case letters, and their alphabets will be denoted
by the respective calligraphic letters.
A similar convention will apply to
random vectors and their sample values
which will be denoted with same symbols in a bold face font.
For example, $y\in\calY$ is a realization of a
random variable $Y$, whereas $\by=(y_1,\ldots,y_n)\in\calY^n$ ($n$ being a
positive integer and $\calY^n$ being the $n$--th Cartesian power of $\calY$)
is a realization of a random vector $\bY=(Y_1,\ldots,Y_n)$.

Let $U$ be a uniformly distributed\footnote{This specific assumption
concerning the density of $U$ and its support is made
for convenience only. Our results extend to more general densities.}
random variable over the interval
$[-1/2,+1/2]$, which we will also denote by $\calU$. 
We refer to $U$ as the parameter to be conveyed from the source
to the destination, via a given noisy channel. A given realization of $U$ will
be denoted by $u$.

A discrete memoryless channel (DMC) is characterized by a matrix of conditional
probabilities $p=\{p(y|x),~x\in\calX,~y\in\calY\}$, where the channel input and output
alphabets, $\calX$ and $\calY$, are assumed finite.\footnote{The finite
alphabet assumption is used mainly for reasons of simplicity. The extension to
continuous alphabets is possible, though some caution should be exercised at
several places.}
When a DMC $p=\{p(y|x),~x\in\calX,~y\in\calY\}$ is fed by
an input vector $\bx\in\calX^n$, it produces an output vector $\by\in\calY^n$
according to
\begin{equation}
p(\by|\bx)=\prod_{t=1}^n p(y_t|x_t).
\end{equation}
A modulator is a measurable mapping $\bx=f_n(u)$ from $\calU=[-1/2,+1/2]$ to
$\calX^n$ and an estimator is a mapping $\hat{u}=g_n(\by)$ from
$\calY^n$ back to $\calU$. The random vector $f_n(U)$ will also be denoted by
$\bX$. Similarly,
the random variable $g_n(\bY)$ will also be
denoted by $\hat{U}$. Our basic figure of merit for communication systems is
the expectation of $\rho$--th power of the estimation error, i.e.,
$\bE\{|\hat{U}-U|^\rho\}$, where $\rho$ is a positive real (not necessarily an
integer) and $\bE\{\cdot\}$ is the expectation operator with respect to
(w.r.t.) the randomness of $U$ and $\bY$. The capability of attaining an
exponential decay in $\bE\{|\hat{U}-U|^\rho\}$ by certain choices of a
modulator $f_n$ and an estimator $g_n$, motivates the definition of
the following exponential rates
\begin{equation}
\overline{\calE}(\rho)=\limsup_{n\to\infty}\left[-\frac{1}{n}\ln\left(\inf_{f_n,g_n}\bE\{|\hat{U}-U|^\rho\}
\right)\right]
\end{equation}
and
\begin{equation}
\underline{\calE}(\rho)=\liminf_{n\to\infty}\left[-\frac{1}{n}\ln\left(\inf_{f_n,g_n}\bE\{|\hat{U}-U|^\rho\}
\right)\right].
\end{equation}
This paper is basically about the derivation of upper bounds on $\overline{\calE}(\rho)$
and lower bounds on $\underline{\calE}(\rho)$, with special interest in
situations where these upper and lower bounds come close to each other.

\section{Upper and Lower Bounds Based on Channel Coding}

Let $q=\{q(x),~x\in\calX\}$ be a given probability vector of a random
variable $X$ taking on values in $\calX$, and let $p=\{p(y|x),~\calX,~y\in\calY\}$
define the given DMC.
Let $E_0(\rho,q)$ be the {\it Gallager function} \cite[p.\ 138, eq.\
(5.6.14)]{Gallager68}, \cite[p.\ 133, eq.\ (3.1.18)]{VO79},
defined as
\begin{equation}
E_0(\rho,q)=-\ln\left(\sum_{y\in\calY}\left[\sum_{x\in\calX}
q(x)p(y|x)^{1/(1+\rho)}\right]^{1+\rho}\right),~~~~\rho\ge 0.
\end{equation}
Next, we define
\begin{equation}
E_0(\rho)=\max_q E_0(\rho,q), 
\end{equation}
where the maximum is over the entire
simplex of probability vectors, and let $\overline{E}_0(\rho)$ be the upper
concave envelope\footnote{While the Gallager function $E_0(\rho,q)$ is known
to be concave in $\rho$ for every fixed $q$ \cite[p.\ 134, eq.\
(3.2.5a)]{VO79}, we are not aware of an argument asserting that $E_0(\rho)$ is
concave in general. On the other hand, there are many situations where
$E_0(\rho)$ is, in fact, concave and then $\overline{E}_0(\rho)=E_0(\rho)$, for example,
when the achiever $q^*$ of $\max_qE_0(\rho,q)$ is independent of $\rho$,
like the case of the binary input output--symmetric (BIOS) channel \cite[p.\
153]{VO79}.} (UCE) of $E_0(\rho)$. Next define
\begin{equation}
E_x(\varrho)=-\varrho\ln\left(\sum_{x,x'\in\calX}
q(x)q(x')\left[\sum_{y\in\calY}\sqrt{p(y|x)p(y|x')}\right]^\varrho\right)
\end{equation}
where the parameter $\varrho$ should be distinguished from the power $\rho$ of
the estimation error in discussion. The {\it expurgated exponent function}
\cite[p.\ 153, eq.\ (5.7.11)]{Gallager68}, \cite[p.\ 146, eq.\ (3.3.13)]{VO79} is
defined as
\begin{equation}
E_{ex}(R)=\sup_{\varrho\ge 1}[E_x(\varrho)-\varrho R].
\end{equation}
It is well known (and a straightforward exercise to show) that
\begin{equation}
E_{ex}(0)=\sup_{\varrho\ge 1}E_x(\varrho)=\lim_{\varrho\to\infty}E_x(\varrho)=
-\sum_{x,x'\in\calX}q(x)q(x')\ln\left[\sum_{y\in\calY}\sqrt{p(y|x)p(y|x')}\right].
\end{equation}
Finally, define
\begin{equation}
\bar{E}(\rho)=\left\{\begin{array}{ll}
\overline{E}_0(\rho) & \rho\le\rho_0\\
E_{ex}(0) & \rho > \rho_0\end{array}\right.
\end{equation}
where $\rho_0$ is the (unique) solution to the equation
$\overline{E}_0(\rho)=E_{ex}(0)$. 

Our first theorem (see Appendix A for the proof) 
asserts that $\overline{E}(\rho)$ is an upper bound on
the best achievable exponential decay rate of $\rho$--th moment of the estimation error.
\begin{theorem}
Let $U$ be uniformly distributed over $\calU=[-1/2,+1/2]$ and let
$p=\{p(y|x)~x\in\calX,~y\in\calY\}$ be a given DMC. Then, for every $\rho\ge 0$
\begin{equation}
\overline{\calE}(\rho)\le \overline{E}(\rho).
\end{equation}
\end{theorem}

We now proceed to present a lower bound $\underline{E}(\rho)$ to
$\underline{\calE}(\rho)$.
Let $R_-$ be the smallest $R$
such that $E_{ex}(R)$ is attained with $\varrho=1$ and let $R_+$ denote the
largest $R$ such that 
\begin{equation}
E_r(R)=\max_{0\le\rho\le 1}[E_0(\varrho,q)-\varrho R]
\end{equation}
is attained for $\varrho=1$.\footnote{For example, in
the case of
the BSC with a crossover parameter $p$, $R_-=\ln 2-h_2(Z/(1+Z))$, with
$Z=\sqrt{4p(1-p)}$, and $R_+=\ln 2-h_2(\sqrt{p}/(\sqrt{p}+\sqrt{1-p}))$, where
$h_2(x)=-x\ln x-(1-x)\ln(1-x)$ \cite[pp.\ 151--152]{VO79}.}
Next, define
\begin{eqnarray}
\rho_+&=&\frac{E_0(1)-R_+}{R_+}\\
\rho_-&=&\frac{E_0(1)-R_-}{R_-}
\end{eqnarray}
and finally,
\begin{equation}
\underline{E}(\rho)=\left\{\begin{array}{ll}
\sup_{0\le \varrho\le 1}\rho E_0(\varrho)/(\varrho+\rho) & \rho \le \rho_+\\
\rho E_0(1)/(1+\rho)=\rho E_x(1)/(1+\rho) & \rho_+< \rho \le \rho_-\\
\sup_{\varrho\ge 1}\rho E_x(\varrho)/(\varrho+\rho) & \rho> \rho_-
\end{array}\right.
\end{equation}
Our next theorem (see Appendix B for the proof) 
tells us that $\underline{E}(\rho)$ is a lower bound on
the best attainable exponential decay rate of $\bE\{|\hat{U}-U|^\rho\}$.
\begin{theorem}
Let $U$ be uniformly distributed over $\calU=[-1/2,+1/2]$ and let
$p=\{p(y|x)~x\in\calX,~y\in\calY\}$ be a given DMC. Then, for every $\rho\ge 0$
\begin{equation}
\underline{\calE}(\rho)\ge \underline{E}(\rho).
\end{equation}
\end{theorem}

The derivations of both $\overline{E}(\rho)$ and $\underline{E}(\rho)$
rely on channel coding considerations. In particular, the derivation of
$\overline{E}(\rho)$ builds strongly on the method of \cite{p152}, which
extends the derivation of the Ziv--Zakai bound \cite{ZZ69} and the Chazan--Zakai--Ziv 
bound \cite{CZZ75}. While the two latter bounds are based on considerations
associated with binary hypotheses testing, here and in \cite{p152}, the
general idea is extended to exponentially many hypotheses pertaining to
channel decoding.

We see that both bounds exhibit
different types of behavior in different ranges of $\rho$ (i.e., ``phase
transitions''), but in
a different manner. For both $\overline{E}(\rho)$ and
$\underline{E}(\rho)$ the behavior is related to the ordinary Gallager
function in some range of small $\rho$, and to the expurgated
exponent in a certain range of large $\rho$. 

As can be seen in the proof of Theorem 2 (Appendix B), the communication
system that achieves $\underline{E}(\rho)$ works as follows (see also
\cite{p152}, \cite{NVW12}): Define
\begin{equation}
R(\rho) = \frac{\underline{E}(\rho)}{\rho}=\left\{\begin{array}{ll}
\sup_{0\le \varrho\le 1}E_0(\varrho)/(\varrho+\rho) & \rho \le
\rho_+\\
E_0(1)/(1+\rho)=E_x(1)/(1+\rho) & \rho_+< \rho \le
\rho_-\\
\sup_{\varrho\ge 1}E_x(\varrho)/(\varrho+\rho) & \rho> \rho_-
\end{array}\right.
\end{equation}
Construct a uniform grid of $M=e^{nR(\rho)}/2$ evenly spaced points along $\calU$,
denoted $\{u_1,u_2,\ldots,u_M\}$. If $\rho > \rho_-$ assign to each grid point $u_i$ a
codeword of a code of rate $R(\rho)$ that achieves the expurgated exponent
$E_{ex}[R(\rho)]$ (see \cite[Theorem 5.7.1]{Gallager68} or \cite[Theorem 3.3.1]{VO79}).
If $\rho \le \rho_-$, do the same with a code that achieves $E_r[R(\rho)]$
(see \cite[p.\ 139, Corollary 1]{Gallager68} or \cite[Theorem 3.2.1]{VO79}).
Given $u$, let $f_n(u)$ be the codeword $\bx_i$ that is assigned to 
the grid point $u_i$, which is closest to $u$. Given $\by$, let $g_n(\by)$ be
the grid point $u_j$ that corresponds to the codeword $\bx_j$ that has been
decoded based on $\by$ using the ML decoder for the given DMC.

Let us examine the behavior of these bounds as $\rho\to 0$ and as
$\rho\to\infty$. For very large values of $\rho)$, where the upper bound $\overline{E}(\rho)$
is obviously given by $E_{ex}(0)$, the lower bound is given by
\begin{eqnarray}
\lim_{\rho\to\infty}\underline{E}(\rho)&=&\lim_{\rho\to\infty}\sup_{\varrho\ge 1}\frac{\rho
E_x(\varrho)}{\varrho+\rho}\\
&\ge&\lim_{\rho\to\infty}\frac{\rho
E_x(\sqrt{\rho})}{\sqrt{\rho}+\rho}\\
&=&\lim_{\rho\to\infty}
E_x(\sqrt{\rho})=E_{ex}(0),
\end{eqnarray}
which means that for large $\rho$ all the exponents asymptotically coincide:
\begin{equation}
\lim_{\rho\to\infty}\underline{E}(\rho)=
\lim_{\rho\to\infty}\underline{\calE}(\rho)=
\lim_{\rho\to\infty}\overline{\calE}(\rho)=
\lim_{\rho\to\infty}\overline{E}(\rho)=E_{ex}(0).
\end{equation}
In the achievability scheme described above, $R(\rho)$ is a very low coding rate.
On the other hand, for very small values of $\rho$, where
$\overline{E}(\rho)=\overline{E}_0(\rho)=\rho C+o(\rho)$, $C$ being the
channel capacity, we have
\begin{eqnarray}
\lim_{\rho\to 0}\frac{\underline{E}(\rho)}{\rho}&=&\lim_{\rho\to 0}\sup_{0\le \varrho\le
1}\frac{
E_0(\varrho)}{\varrho+\rho}\\
&\ge&\lim_{\rho\to 0}\frac{
E_0(\sqrt{\rho})}{\sqrt{\rho}+\rho}\\
&=&\lim_{\rho\to 0}
\frac{E_0(\sqrt{\rho})}{\sqrt{\rho}}\cdot\frac{1}{1+\sqrt{\rho}}\\
&=&\lim_{\rho\to 0}
\frac{E_0(\sqrt{\rho})}{\sqrt{\rho}}=C,
\end{eqnarray}
which means that for small $\rho$ all the exponents behave like $\rho C$,
i.e.,
\begin{equation}
\lim_{\rho\to 0}\frac{\underline{E}(\rho)}{\rho}=
\lim_{\rho\to 0}\frac{\underline{\calE}(\rho)}{\rho}=
\lim_{\rho\to 0}\frac{\overline{\calE}(\rho)}{\rho}=
\lim_{\rho\to 0}\frac{\overline{E}(\rho)}{\rho}=C.
\end{equation}
It is then interesting to observe that not only channel--coding error
exponents, but also channel capacity
plays a role in the characterization of the best achievable
modulation--estimation performance.
In the achievability scheme described above, 
$R(\rho)$ is a very high coding rate,
very close to the capacity $C$.

\section{Upper Bound Based on Data Processing
Inequalities}

We next derive an alternative upper bound on $\overline{\calE}(\rho)$ that is
based on generalized data processing inequalities, following
Ziv and Zakai \cite{ZZ73} and Zakai and Ziv \cite{ZZ75}. The idea behind these
works is that it is possible to define generalized mutual information
functionals satisfying a DPT, by replacing the negative logarithm function of the ordinary
mutual information, by a general convex
function. This enables to obtain tighter distortion bounds for communication
systems with short block length.

In \cite{p146} it was shown that
the following generalized mutual information functional,
between two generic random variables, $A$ and $B$, admits a DPT for every positive integer $k$ and for
every vector $(\alpha_1,\ldots,\alpha_k)$ whose components are non--negative
and sum to unity:
\begin{equation}
\tilde{I}(A;B)=-\bE\left\{\sum_{b\in\calB}\prod_{i=1}^k
p(b|A_i)^{\alpha_i}\right\}=-\sum_{b\in\calB}\prod_{i=1}^k\sum_{a_i\in\calA}
q(a_i)p(b|a_i)^{\alpha_i}.
\end{equation}
In particular,
since $U\to \bY\to \hat{U}$ is a Markov chain, then by the generalized DPT,
\begin{equation}
\tilde{I}(U;\hat{U})\le \tilde{I}(U;\bY). 
\end{equation}
The idea is to further upper bound
$I(U;\bY)$ and to further lower bound $\tilde{I}(U;\hat{U})$ subject to the
constraint $\bE|\hat{U}-U|^\rho = D$, which leads to a generalized
rate--distortion function,
and thereby to obtain an inequality on $\bE|\hat{U}-U|^\rho$.
Specifically, $I(U;\bY)$ is upper bounded as follows:
\begin{eqnarray}
\tilde{I}(U;\bY)&=&-\sum_{\by\in\calY^n}\prod_{i=1}^k\int_{-1/2}^{+1/2}
\mbox{d}u_i p(\by|f_n(u_i))^{\alpha_i}\\
&=&-\sum_{\by\in\calY^n}\prod_{i=1}^k\int_{-1/2}^{+1/2}
\mbox{d}u_i \prod_{t=1}^np(y_t|[f_n(u_i)]_t)^{\alpha_i}\\
&=&-\prod_{t=1}^n\sum_{y\in\calY}\prod_{i=1}^k\int_{-1/2}^{+1/2}
\mbox{d}u_ip(y_t|[f_n(u_i)]_t)^{\alpha_i}\\
&\le&-\min_q\prod_{t=1}^n\sum_{y\in\calY}\prod_{i=1}^k\sum_{x_i\in\calX}q(x_i)
p(y_t|x_i)^{\alpha_i}\\
&=&-\min_q\left[\sum_{y\in\calY}\prod_{i=1}^k\sum_{x_i\in\calX}q(x_i)
p(y|x_i)^{\alpha_i}\right]^n\\
&=&-\exp\{-n\max_qE(\alpha_1,\ldots,\alpha_k,q)\},
\end{eqnarray}
where $[f_n(u_i)]_t$ denotes the $t$--th component of the vector $\bx=f_n(u_i)$
and where
\begin{equation}
E(\alpha_1,\ldots,\alpha_k,q)=-\ln\left[\sum_{y\in\calY}\prod_{i=1}^k\left(\sum_{x_i\in\calX}
q(x_i)p(y|x_i)^{\alpha_i}\right)\right].
\end{equation}
Note that for $k=1+\varrho$ ($\varrho$ -- integer),
\begin{equation}
\hat{E}\left(\frac{1}{1+\varrho},\ldots,\frac{1}{1+\varrho},q\right)=E_0(\varrho,q).
\end{equation}
In Appendix C we show that
\begin{equation}
\label{rd}
\min\{\tilde{I}(U,\hat{U}):~\bE|\hat{U}-U|^\rho = D\}\dfn
\tilde{R}(D)\ge - c\cdot D^{\sum_{i=1}^k\zeta_\rho(\alpha_i)}
\end{equation}
where $c$ is a constant that depends solely on $\rho$, $k$ and
$\alpha_1,\ldots,\alpha_k$, and where
\begin{equation}
\label{zeta}
\zeta_\rho(\alpha)=\left\{\begin{array}{ll}
\alpha & 0\le \alpha \le \frac{1}{1+\rho}\\
\frac{1-\alpha}{\rho} & \frac{1}{1+\rho} \le \alpha \le 1\end{array}\right.
=\min\left\{\alpha,\frac{1-\alpha}{\rho}\right\}.
\end{equation}
The function $\tilde{R}(D)$ in eq.\ (\ref{rd}) is referred to as a ``generalized rate--distortion
function'' in the terminology of \cite{ZZ73} and \cite{ZZ75}.
Thus, from the generalized DPT,
\begin{equation}
\bE|\hat{U}-U|^\rho\equiv D \ge c'\cdot e^{-n\overline{E}_{DPT}(\rho)}
\end{equation}
where $c'$ is another constant and
\begin{equation}
\overline{E}_{DPT}(\rho)\dfn\inf_{k>1}\inf_{\alpha_1,\ldots,\alpha_k}\sup_q
\frac{E(\alpha_1,\ldots,\alpha_k,q)}{\sum_{i=1}^k\zeta_\rho(\alpha_i)}.
\end{equation}
As an example, assume that the channel is such that the function $E_0(\varrho)$ is
concave, so that $\overline{E}_0(\varrho)=E_0(\varrho)$. In this case,
$\rho_0\ge 1$ since $E_0(1) \le E_{ex}(0)$ and $E_0(\varrho)$ is monotonically
increasing. Now,
let $\rho \le \rho_0$ be an integer (for example, $\rho=1$ is always a
legitimate choice). Then,
\begin{eqnarray}
\overline{E}(\rho)
&=&E_0(\rho)\\
&=&\sup_q\frac{E(1/(1+\rho),\ldots,1/(1+\rho),q)}{(1+\rho)\zeta_\rho(1/(1+\rho))}\\
&\ge&
\inf_{k>1}\inf_{\alpha_1,\ldots,\alpha_k}\sup_q\frac{\hat{E}(\alpha_1,\ldots,\alpha_k,q)}
{\sum_{i=1}^k\zeta_\rho(\alpha_i)}\\
&=&\overline{E}_{DPT}(\rho).
\end{eqnarray}
Thus, at least in this case, the DPT bound is 
guaranteed to be no worse than the channel--coding bound $\overline{E}(\rho)$.
Nonetheless, in our numerical studies, we have not found an example where
the DPT bound strictly improves on the channel--coding bound, i.e.,
$\overline{E}_{DPT}(\rho) < \overline{E}(\rho)$, and it remains an open
question whether the DPT bound can offer improvement in any situation, thanks
to its additional degrees of freedom. It should be pointed out that the vector
$(\alpha_1,\ldots,\alpha_k)$ that achieves $E_{DPT}(\rho)$ is not always
given by $(1/(k+1),\ldots,1/(k+1))$ because the function
$E(\alpha_1.\ldots,\alpha_k,q)$ is not convex in $(\alpha_1,\ldots,\alpha_k)$.
At any rate, in all cases where the two bounds are equivalent, namely, $\overline{E}_{DPT}(\rho)=
\overline{E}(\rho)$, this is interesting on its own right
since the two bounds are obtained by two different techniques that are based on 
completely different considerations. One advantage of the DPT approach is
that it seems to lend itself more comfortably to extensions that account
for moments of more general functions of the estimation error, i.e.,
$\bE\{g(|\hat{U}-U|)\}$, for a large class of monotonically increasing
functions $g$. On the other hand, the optimization associated with calculation
of the DPT bound is not trivial.

\section{Example: Very Noisy Channel}

\begin{figure}[h!t!b!]
\centering
\includegraphics[width=8.5cm, height=8.5cm]{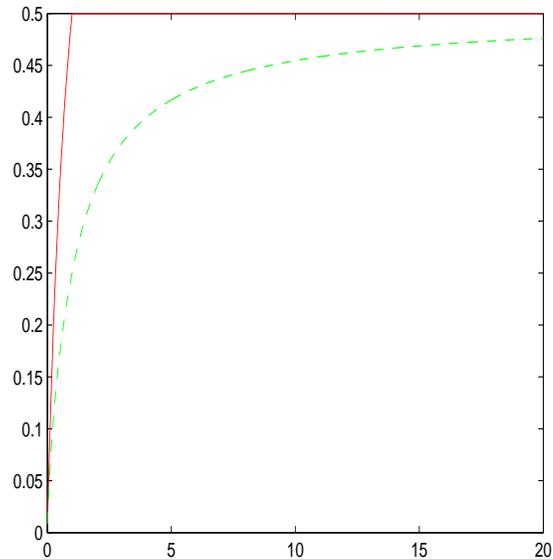}
\caption{The upper bound $\overline{E}(\rho)/C$ (solid curve)
and the lower bound $\underline{E}(\rho)/C$ (dashed curve) for the example of
the very noisy
channel.}
\label{bounds}
\end{figure}

As an example, we consider the so called {\it very noisy channel}, which is
characterized by
\begin{equation}
p(y|x)=p(y)[1+\epsilon(x,y)],~~~~~~|\epsilon(x,y)| \ll 1,~~~\forall
x\in\calX,~y\in\calY.
\end{equation}
As is shown in \cite[Sect.\ pp.\ 155--158]{VO79}, to the first order, we have the following relations
\begin{equation}
C=\frac{1}{2}\max_q\sum_{x,y}q(x)p(y)\epsilon^2(x,y)
\end{equation}
\begin{equation}
E_0(\varrho)=\frac{\varrho}{1+\varrho}\cdot C,
\end{equation}
and therefore
\begin{equation}
E_r(R)=\max_{0\le\varrho\le 1}\left(\frac{\varrho}{1+\varrho}\cdot C-\varrho R\right)=
\left\{\begin{array}{ll}
\frac{C}{2}-R & R <\frac{C}{4}\\
(\sqrt{C}-\sqrt{R})^2 & \frac{C}{4}\le R\le C\\
0 & R > C\end{array}\right.
\end{equation}
As for the expurgated exponent, we have
\begin{equation}
E_x(\varrho)=E_0(1)=\frac{C}{2}
\end{equation}
and so,
\begin{equation}
E_{ex}(R)=\sup_{\varrho\ge 1}[E_x(\varrho)-\varrho R]=\frac{C}{2}-R
\end{equation}
which means that expurgation does not help for very noisy channels.
This implies that $\rho_0=1$ and so
\begin{equation}
\overline{E}(\rho)=\left\{\begin{array}{ll}
\frac{\rho}{1+\rho}\cdot C & \rho\le 1\\
\frac{C}{2} & \rho > 1\end{array}\right.
\end{equation}
As for the lower bound, we have the following:
For $\rho< 1$, 
\begin{equation}
\underline{E}(\rho)=\sup_{0\le \varrho\le
1}\frac{\rho}{\rho+\varrho}\cdot\frac{\varrho}{1+\varrho}\cdot
C=\frac{\rho}{(1+\sqrt{\rho})^2}\cdot C.
\end{equation}
The same result is obtained, of course, from the solution to the equation
$\rho R=(\sqrt{C}-\sqrt{R})^2$. For $\rho\ge 1$, 
\begin{equation}
\underline{E}(\rho)=\sup_{\varrho\ge 1}\frac{\rho E_x(\varrho)}{\varrho+\rho}=
\sup_{\varrho\ge
1}\frac{\rho}{\varrho+\rho}\cdot\frac{C}{2}=\frac{\rho}{1+\rho}\cdot\frac{C}{2}.
\end{equation}
Thus, in summary
\begin{equation}
\underline{E}(\rho)=\left\{\begin{array}{ll}
\frac{\rho}{(1+\sqrt{\rho})^2}\cdot C & \rho < 1\\
\frac{\rho}{1+\rho}\cdot\frac{C}{2} & \rho \ge 1
\end{array}\right.
\end{equation}
We see how the bounds asymptotically coincide (in the sense that 
$\overline{E}(\rho)/\underline{E}(\rho)\approx 1$) both for very large values
of $\rho$ and for 
very small values of $\rho$ (see Fig.\ \ref{bounds}).

As for the DPT bound, we have the following approximate analysis:
\begin{eqnarray}
e^{-\sup_qE(\alpha_1,\ldots,\alpha_k,q)}&=&
\inf_q\sum_{y\in\calY}\prod_{i=1}^k\left[\sum_{x_i\in\calX}q(x_i)p(y|x_i)^{\alpha_i}\right]\\
&=&\inf_q\sum_{y\in\calY}\prod_{i=1}^k\left\{p(y)^{\alpha_i}
\left[\sum_{x_i\in\calX}q(x_i)[1+\epsilon(x_i,y)]^{\alpha_i}\right]\right\}\\
&=&\inf_q\sum_{y\in\calY}p(y)\prod_{i=1}^k\left[\sum_{x_i\in\calX}
q(x_i)[1+\epsilon(x_i,y)]^{\alpha_i}\right]\\
&\approx&\inf_q
\sum_{y\in\calY}p(y)\prod_{i=1}^k\left(\sum_{x_i\in\calX}q(x_i)\left[1+\alpha_i\epsilon(x_i,y)-
\frac{1}{2}\alpha_i(1-\alpha_i)\epsilon^2(x_i,y)\right]\right)\\
&=&\inf_q\sum_{y\in\calY}p(y)\prod_{i=1}^k\left[1-
\frac{1}{2}\alpha_i(1-\alpha_i)\sum_{x_i\in\calX}q(x_i)\epsilon^2(x_i,y)\right]\\
&\approx&\inf_q\sum_{y\in\calY}p(y)\left[1-
\frac{1}{2}\sum_{i=1}^k\alpha_i(1-\alpha_i)\sum_{x_i\in\calX}q(x_i)\epsilon^2(x_i,y)\right]\\
&=&1-
\frac{1}{2}\sum_{i=1}^k\alpha_i(1-\alpha_i)\sup_q\sum_{x_i\in\calX}
\sum_{y\in\calY}q(x_i)p(y)\epsilon^2(x_i,y)\\
&\approx& 1-C\sum_{i=1}^k\alpha_i(1-\alpha_i)\\
&=& 1-C\left(1-\sum_{i=1}^k\alpha_i^2\right).
\end{eqnarray}
where in the fifth line,
we have used the identity $\sum_{x}q(x)\epsilon(x,y)=0$ for all $y$ with
$p(y)>0$ \cite[p.\
156, eq.\ (3.4.28)]{VO79}.
Thus,
\begin{equation}
\sup_qE(\alpha_1,\ldots,\alpha_k,q)=
-\ln\left[1-C\left(1-\sum_{i=1}^k\alpha_i^2\right)\right]
\approx C\left(1-\sum_{i=1}^k\alpha_i^2\right),
\end{equation}
and then
\begin{equation}
\overline{E}_{DPT}(\rho)\approx C\cdot \inf_{k>1}\inf_{\alpha_1,\ldots,\alpha_k}
\frac{1-\sum_{i=1}^k\alpha_i^2}{\sum_{i=1}^k
\zeta_\rho(\alpha_i)}.
\end{equation}
The very same expressions are obtained for the continuous--time AWGN channel with
unlimited bandwidth, where $C=P/N_0$, $P$ being the signal power and $N_0$
being the one--sided noise spectral density.
For $\rho=1$ and $k=2$, we have $\zeta_1(\alpha)=\min\{\alpha,1-\alpha\}$:
\begin{eqnarray}
\overline{E}_{DPT}(1)&\le&C\cdot\inf_{0\le \alpha\le 1}
\frac{1-\alpha^2-(1-\alpha)^2}{2\min\{\alpha,1-\alpha\}}\\
&=&C\cdot\inf_{0\le\alpha\le 1/2}\frac{2\alpha(1-\alpha)}{2\alpha}
=\frac{C}{2},
\end{eqnarray}
which agrees with $\overline{E}(\rho)$.
For $\rho=2$ and $k=2$, the minimum is attained for $\alpha=1/3$, and the result
is $E_{DPT}(2)\le 8C/9$. However for $k=3$, the bound improves to $C/3$.

\section{Extension to the Multidimensional Case}

Consider now the case of a parameter vector $\bU=(U_1,\ldots, U_d)$, uniformly
distributed across the unit hypercube $[-1,2,+1/2]^d$. A reasonable figure of merit 
in this case would be
a linear combination of $\bE\{|\hat{U}_i-U_i|^\rho\}$, $i=1,2,\ldots,d$. Since
each one of these terms is exponential in $n$, it makes sense to let the
coefficients of this linear combination also be exponential functions of $n$,
as otherwise, the results will be exponentially insensitive to the choice of the
coefficients. This means that we consider the criterion
\begin{equation}
\sum_{i=1}^d e^{nr_i}\cdot\bE\{|\hat{U}_i-U_i|^\rho\},
\end{equation}
where, without loss of generality, we take $r_i\ge 0$, $\min_ir_i=0$.

The derivation below is an extension of the derivation of the channel coding
bound, given in Appendix A for the case $d=1$. Therefore, a reader who is
interested in the details is advised to read Appendix A
first, or otherwise to skip directly to the final result in eq.\
(\ref{endresult}) and the discussion that follows.

Let us define $R_i=(r_i+\gamma)/\rho$ for some constant $\gamma\ge 0$.
Consider the following chain of inequalities:
\begin{eqnarray}
\sum_{i=1}^d e^{nr_i}\cdot\bE\{|\hat{U}_i-U_i|^\rho\}
&\ge&\sum_{i=1}^d e^{nr_i}\cdot e^{-n\rho R_i}\mbox{Pr}\{|\hat{U}_i-U_i|\ge
e^{-nR_i}\}\\
&=&\sum_{i=1}^d e^{-n(\rho R_i-r_i)}\mbox{Pr}\{|\hat{U}_i-U_i|\ge
e^{-nR_i}\}\\
&=&e^{-\gamma n}\sum_{i=1}^d \mbox{Pr}\{|\hat{U}_i-U_i|\ge
e^{-n(r_i+\gamma)/\rho}\}\\
&\ge&e^{-\gamma n}\cdot\mbox{Pr}\bigcup_{i=1}^d\left\{|\hat{U}_i-U_i|\ge
e^{-n(r_i+\gamma)/\rho}\right\}\\
&\ge&e^{-\gamma n}\cdot
\exp\left\{-nE_{sl}\left(\frac{1}{\rho}\left[\sum_{i=1}^dr_i+\gamma
d\right]\right)\right\},
\end{eqnarray}
where the second line follows from Chebychev's inequality, the fifth line
follows from the union bound, and
the last line follows from the same arguments as in \cite[Sect.\
IV.A]{p152}.
Maximizing over $\gamma$, we get
\begin{equation}
\sum_{i=1}^d e^{nr_i}\cdot\bE\{|\hat{U}_i-U_i|^\rho\}\\
\ge\exp\left\{-n\min_{\gamma\ge
0}\left[\gamma+E_{sl}\left(\frac{1}{\rho}\left[\sum_{i=1}^dr_i+\gamma
d\right]\right)\right]\right\}.
\end{equation}
Defining $R=(\sum_{i=1}^d r_i+\gamma d)/\rho$,
$R_{\min}=\sum_{i=1}^dr_i/\rho$ and
$\bar{r}=R_{\min}/d$,
the above minimization at the exponent becomes equivalent to
\begin{eqnarray}
& &\min_{R\ge R_{\min}}\left[\frac{\rho R-\sum_ir_i}{d}+E_{sl}(R)\right]\\
&=&\min_{R\ge R_{\min}}\left[\frac{\rho}{d}\cdot
R+E_{sl}(R)\right]-\rho\bar{r}\\
&=&\left\{\begin{array}{ll}
E_{sp}(R_{\rho/d})+\frac{\rho}{d}(R_{\rho/d}-R_{\min}) & \rho/d \le \rho_0\\
E_{ex}(0)-\rho_0 R_{\min} & \rho/d > \rho_0
\end{array}\right.
\end{eqnarray}
where $R_\theta$ is defined as the achiever of $\min_{R\ge R_{\min}}[\theta
R+E_{sp}(R)]$. Thus, the extension of the channel--coding bound to the
$d$--dimensional case reads
\begin{eqnarray}
\overline{E}(\rho,d,r_1,\ldots,r_d)&=&
\left\{\begin{array}{ll}
E_{sp}(R_{\rho/d})+\frac{\rho}{d}R_{\rho/d}-\frac{1}{d}\sum_{i=1}^dr_i &
\rho\le \rho_0d\\
E_{ex}(0)-\frac{\rho_0}{\rho}\sum_{i=1}^dr_i & \rho > \rho_0d
\end{array}\right.\\
&=&\left\{\begin{array}{ll}
E_0\left(\frac{\rho}{d}\right)-\frac{1}{d}\sum_{i=1}^dr_i &
\rho/d\le \rho_0\\
E_{ex}(0)-\frac{\rho_0}{\rho}\sum_{i=1}^dr_i & \rho/d > \rho_0
\end{array}\right.
\label{endresult}
\end{eqnarray}
We see that when $r_i=0$ for all $i$ (i.e., all weights are 1), it is the same
channel--coding bound
as before, except that $\rho$ is replaced by $\rho/d$, that is,
$\overline{E}(\rho/d)$.
For $\rho\to\infty$, the
bound tends to $E_{ex}(0)$, which can be
approached again by a low--rate code for a
Cartesian grid in the parameter space.
At the other extreme, when $d$ is very large compared to $\rho$, so $\rho/d$
is small, construct a grid of $e^{n(C-\epsilon)/d}\times
e^{n(C-\epsilon)/d}\times \ldots \times e^{n(C-\epsilon)/d}$, quantize $\bU$
and assign to each grid point a codeword of a typical random code at rate
$C-\epsilon$. Then the performance will be about $e^{-n\rho C/d}$.
Therefore, as a corollary of the above result, we have
\begin{equation}
\sum_{i=1}^d\bE\{|\hat{U}_i-U_i|^\rho\}\\
\ge e^{-n[\overline{E}(\rho/d)+o(n)]}.
\end{equation}

\section*{Appendix A}
\renewcommand{\theequation}{A.\arabic{equation}}
    \setcounter{equation}{0}
\noindent
{\it Proof of Theorem 1.}
We begin by using
the Markov/Chebychev inequality:
\begin{equation}
\bE|\hat{U}-U|^\rho\ge \Delta^\rho\mbox{Pr}\{|\hat{U}-U|\ge \Delta\}.
\end{equation}
Next we need to further lower bound $\mbox{Pr}\{|\hat{U}-U|\ge \Delta\}$ and then
maximize the r.h.s.\ over $\Delta$. Equivalently, similarly as in \cite{p152}, we may set
$\Delta=e^{-nR}$
in the r.h.s.\ and maximize the bound w.r.t.\ $R$. Let $E(R)$ be the
reliability function of the channel. Then, similarly\footnote{While ref.\
\cite{p152} is primarily about the continuous time additive white Gaussian
noise (AWGN) channel, the arguments in the proof of Theorem 1 therein are insensitive
to this assumption. They hold verbatim here, provided that the observation time
$T$ in \cite{p152} is replaced by the block length $n$ and the reliability function of the AWGN
channel is replaced by that of the DMC considered here.}
as in \cite[Theorem
1]{p152}, we have:
\begin{equation}
\mbox{Pr}\{|\hat{U}-U|\ge e^{-nR}\}\ge e^{-n[E(R)+o(n)]}
\end{equation}
and so,
\begin{equation}
\bE|\hat{U}-U|^\rho\ge e^{-n\rho R}\cdot e^{-n[E(R)+o(n)]}=e^{-n[\rho
R+E(R)+o(n)]}.
\end{equation}
The best\footnote{The reader might suspect that the use of Chebychev's
inequality yields a loose bound. Note, however, that even the exact relation
$\bE|\hat{U}-U|^\rho=\rho n\int_0^\infty\mbox{d}R\cdot e^{-n\rho
R}\cdot\mbox{Pr}\{|\hat{U}-U|> e^{-nR}\}$, with $\mbox{Pr}\{|\hat{U}-U|>
e^{-nR}\}\ge e^{-n[E(R)+o(n)]}$, would yield, after saddle--point integration, exactly the same exponential
order as presented above. 
The weak link here is, therefore, not the Chebychev inequality but the fact that
there is no apparent single estimator, independent of $R$, that minimizes
$\mbox{Pr}\{|\hat{U}-U|>e^{-nR}\}$ uniformly for all $R$.}
lower bound is obtained by maximizing the r.h.s.\ over $R$,
yielding
\begin{eqnarray}
\label{lb}
\bE|\hat{U}-U|^\rho&\ge&e^{-n\min_{R\ge 0}[\rho R+E(R)+o(n)]}\nonumber\\
&\ge&e^{-n\min_{R\ge 0}[\rho R+E_{sl}(R)+o(n)]}
\end{eqnarray}
where $E_{sl}(R)$ is the exponent associated with the {\it straight line
bound}, which is well known to be an upper bound on the reliability function
$E(R)$ \cite{SGB67a}, \cite{SGB67b}, 
\cite[Sect.\ 3.8]{VO79}, and which is given by
\begin{equation}
E_{sl}(R)=\left\{\begin{array}{ll}
E_{ex}(0)-\rho_0R & 0\le R \le R_0\\
E_{sp}(R) & R_0 < R \le C\\
0 & R > C\end{array}\right.
\end{equation}
where
\begin{equation}
\label{Esp}
E_{sp}(R)=\sup_{\varrho\ge 0}[E_0(\varrho)-\varrho R]
\end{equation}
is the {\it sphere--packing exponent}, $\rho_0$ is as defined in Theorem 1 and
$R_0$ is the rate $R$ at which $\mbox{d}E_{sp}(R)/\mbox{d}R=-\rho_0$, or
equivalently, the solution to the equation $E_{sp}(R)=E_{ex}(0)-\rho_0R$.
Thus, according to the second line of eq.\ (\ref{lb}),
\begin{equation}
\overline{\calE}(\rho)\le\min_{R\ge 0}[\rho
R+E_{sl}(R)].
\end{equation}
For $\rho\ge \rho_0$, the minimum is obviously 
attained at $R=0$, and so,
\begin{equation}
\overline{\calE}(\rho)\le \rho\cdot 0+E_{sl}(0)=E_{ex}(0).
\end{equation}
For $\rho< \rho_0$, we use
\begin{equation}
\label{last}
\overline{\calE}(\rho)\le\min_{R\ge 0}[\rho
R+E_{sl}(R)]\le
\min_{R\ge 0}[\rho
R+E_{sp}(R)].
\end{equation}
The right--most side of eq.\ (\ref{last}) is the Legendre--Fenchel transform (LFT) of $E_{sp}(R)$,
which in turn (according to (\ref{Esp})), is the LFT of $E_0(\rho)$. Thus, the
right--most side of (\ref{last}) is given by the UCE of $E_0(\rho)$, which is
$\overline{E}_0(\rho)$. Thus,
\begin{equation}
\overline{\calE}(\rho)\le\left\{\begin{array}{ll}
\overline{E}_0(\rho) & \rho<\rho_0\\
E_{ex}(0) & \rho\ge \rho_0\end{array}\right. =\overline{E}(\rho).
\end{equation}
This completes the proof of Theorem 1.

\section*{Appendix B}
\renewcommand{\theequation}{B.\arabic{equation}}
    \setcounter{equation}{0}
\noindent{\it Proof of Theorem 2.}
Define
\begin{equation}
R(\rho) = \frac{\underline{E}(\rho)}{\rho}=\left\{\begin{array}{ll}
\sup_{0\le \varrho\le 1}E_0(\varrho)/(\varrho+\rho) & \rho \le
\rho_+\\
E_0(1)/(1+\rho)=E_x(1)/(1+\rho) & \rho_+< \rho \le
\rho_-\\
\sup_{\varrho\ge 1}E_x(\varrho)/(\varrho+\rho) & \rho> \rho_-
\end{array}\right.
\end{equation}
Consider a grid of $M=e^{nR(\rho)}/2$ evenly spaced points along $\calU$,
denoted $\{u_1,u_2,\ldots,u_M\}$, where $u_1=-1/2+e^{-nR(\rho)}$ and
$u_M=1/2-e^{-nR(\rho)}$
(see also \cite[Theorem 2]{p152}). 
If $\rho > \rho_-$, assign to each point $u_i$ a
codeword of a code of rate $R(\rho)$ that achieves the expurgated exponent
$E_{ex}[R(\rho)]$. Otherwise,
do the same with a code that achieves $E_r[R(\rho)]$
(see \cite[p.\ 139, Corollary 1]{Gallager68} or \cite[Theorem 3.2.1]{VO79}).
Given $u$, let $f_n(u)$ be the codeword $\bx_i$ that is assigned to
the grid point $u_i$, which is closest to $u$. Given $\by$, let $g_n(\by)$ be
the grid point $u_j$ that corresponds to the codeword $\bx_j$ that has been
decoded based on $\by$ using the ML decoder for the given DMC.
For every $R\ge 0$, we have:
\begin{eqnarray}
\label{ub}
\bE\{|\hat{U}-U|^\rho\}&=&
\bE\left\{|\hat{U}-U|^\rho\bigg| |\hat{U}-U|\le
e^{-nR}\right\}\cdot\mbox{Pr}\{|\hat{U}-U|\le e^{-nR}\}+\nonumber\\
& &\bE\left\{|\hat{U}-U|^\rho\bigg| |\hat{U}-U|>
e^{-nR}\right\}\cdot\mbox{Pr}\{|\hat{U}-U|> e^{-nR}\}\nonumber\\
&\le& [e^{-nR}]^\rho\cdot 1+1^\rho\cdot
\mbox{Pr}\{|\hat{U}-U|> e^{-nR}\}\nonumber\\
&=& e^{-n\rho R}+
\mbox{Pr}\{|\hat{U}-U|> e^{-nR}\}.
\end{eqnarray}
Now, it follows from the construction of the proposed scheme that if $R$ is
the coding rate and the spacing between each two consecutive grid points is
$2e^{-nR}$, then the
event $\{|\hat{U}-U|> e^{-nR}\}$ occurs iff the ML decoder errs. Thus,
$\mbox{Pr}\{|\hat{U}-U|> e^{-nR}\}$
is exactly the probability of decoding error. Considering the case $\rho >
\rho_-$, this code
is assumed to achieve the expurgated exponent, and so, this probability of error is
upper bounded by $e^{-n\{E_{ex}(R)]-o(n)\}}$.
Since $\rho R$ is an increasing function of $R$ and $E_{ex}(R)$ is a
decreasing function, the best choice of $R$ is the solution to the
equation
\begin{equation}
\rho R=E_{ex}(R)
\end{equation}
or, equivalently
\begin{equation}
\label{equ}
\rho R=\sup_{\varrho\ge 1}[E_x(\varrho)-\varrho R].
\end{equation}
Below we show that the solution to this equation is
given by
\begin{equation}
\label{sol}
R=R(\rho)\dfn\sup_{\varrho\ge 1}\frac{E_x(\varrho)}{\varrho+\rho}
\end{equation}
and for this choice of $R$, both exponents in the last line of (\ref{ub})
are given by
\begin{equation}
\rho R(\rho)=\sup_{\varrho\ge 1}\frac{\rho E_x(\varrho)}{\varrho+\rho}
\end{equation}
which is exactly the expression of $\underline{E}(\rho)$ in the range
$\rho > \rho_-$. In the range $\rho < \rho_+$, exactly the same arguments
hold, except that $E_{ex}(R)$ and $E_x(\varrho)$ and
$\sup_{\varrho\ge 1}$ are replaced by
$E_r(R)$, $E_0(\varrho)$, and
$\sup_{0\le\varrho\le 1}$, respectively. In the intermediate range,
the same line of arguments hold once again, with 
$\varrho=1$ and $E_x(1)\equiv E_0(1)$.

It remains to show that $R(\rho)$ in (\ref{sol}) solves equation
(\ref{equ}) for $\rho > \rho_-$, and then similar arguments will follow for the two
other ranges. Let $R(\rho)$ be defined as in (\ref{sol}) and let
$R'(\rho)$ be defined as the solution to (\ref{equ}). We wish to prove that
$R(\rho)=R'(\rho)$. To this end, we will prove that both
$R(\rho)\ge R'(\rho)$ and
$R(\rho)\le R'(\rho)$.
To prove the first inequality, let $\varrho(R)$ denote the
achiever of $E_{ex}(R)=\sup_{\varrho\ge 1}[E_x(\varrho)-\varrho 
R]$. Then, by definition of
$R'(\rho)$, we obviously have
\begin{equation}
\rho R'(\rho)=E_x[\varrho(R'(\rho))]-\varrho[R'(\rho)]R'(\rho)
\end{equation}
i.e.,
\begin{equation}
R'(\rho)=\frac{E_x[\varrho(R'(\rho))]}{\varrho[R'(\rho)]+\rho}\le
\sup_{\varrho\ge
1}\frac{E_x(\varrho)}{\varrho+\rho}\equiv R(\rho).
\end{equation}
To prove the second (opposite) inequality, let $\varrho(\rho)$ be the achiever of
$R(\rho)$, that is,
\begin{equation}
R(\rho)=\frac{E_x[\varrho(\rho)]}{\varrho(\rho)+\rho},
\end{equation}
or, equivalently,
\begin{equation}
\rho R(\rho)=E_x[\varrho(\rho)]-\varrho(\rho)R(\rho).
\end{equation}
But the l.h.s.\ cannot exceed $\sup_{\varrho\ge
1}[E_x(\varrho)-\varrho R(\rho)]=E_{ex}[R(\rho)]$, and so,
\begin{equation}
\rho R(\rho)\le E_{ex}[R(\rho)].
\end{equation}
Now, as mentioned earlier, the function $\rho R$ is increasing in $R$ whereas the function
$E_{ex}(R)$ is decreasing. Thus, the value of $R$ for which there is
equality $\rho R=E_{ex}(R)$, which is $R'(\rho)$, cannot be smaller than any
value of $R$,
for which $\rho R\le E_{ex}(R)$, like $R(\rho)$. Hence, $R(\rho)\le
R'(\rho)$. This completex the proof of Theorem 2.

\section*{Appendix C}
\renewcommand{\theequation}{C.\arabic{equation}}
    \setcounter{equation}{0}

{\it Derivation of a lower bound on the generalized rate--distortion function.}
Consider the minimization of the generalized mutual information
\begin{equation}
\tilde{I}(U;\hat{U})=-
\bE\left\{\int_{\calU}\mbox{d}\hat{u}\prod_{i=1}^k
p(\hat{u}|U_i)^{\alpha_i}\right\}=-\int_{\calU}\mbox{d}\hat{u}\prod_{i=1}^k\int_{\calU}\mbox{d}u_i
p(u_i)p(\hat{u}|u_i)^{\alpha_i}.
\end{equation}
Similarly as in \cite[Sect.\ IV, Example 2]{ZZ73} and \cite{p146}, 
since we are dealing with an exponentially small estimation error level
(small distortion), then
for reasons of convenience, we approximate our distortion measure
$d(u,\hat{u})=|\hat{u}-u|^\rho$ ($u,~\hat{u}\in\calU$) by
\begin{equation}
d'(u,\hat{u})=|(\hat{u}-u)~\mbox{mod}~1|^\rho.
\end{equation}
where
\begin{equation}
t~\mbox{mod}~1\dfn \left< t+\frac{1}{2}\right>-\frac{1}{2}
\end{equation}
$\left<r\right>$ being the fractional part of $r$,
that is, $\left<r\right>=r-\lfloor r\rfloor$.
The justification is that for very small distortion (the high--resolution
limit), the modulo 1 operation has a
negligible effect, and hence
$d'(u,\hat{u})$ becomes essentially equivalent to the original
distortion measure $d(u,\hat{u})=|\hat{u}-u|^\rho$. 
Using the same reasoning as in \cite[Sect.\ IV, Example 2]{ZZ73} and
\cite{p146}, there is no loss of optimality by
confining attention to channels $p(\hat{u}|u)$ of the form $f(w)$
with $w=\hat{u}-u~\mbox{mod}~1$. Thus, the minimization of $\tilde{I}(U;\hat{U})$
reduces to the maximization of
\begin{equation}
U(f)=\prod_{i=1}^k\mbox{d}w_i\int_{-1/2}^{+1/2}\mbox{d}w_i
[f(w_i)]^{\alpha_i}
\end{equation}
subject to the constraints
\begin{eqnarray}
\int_{-1/2}^{+1/2}\mbox{d}w\cdot f(w)&=&1\\
\int_{-1/2}^{+1/2}\mbox{d}w\cdot |w|^\rho f(w)&=& D.
\end{eqnarray}
This optimization problem is not trivial, but we can find an upper bound on
$U(f)$ in terms of $D$ for small $D$.
We begin with the following bound for each one of the factors of $U(f)$:
\begin{eqnarray}
\int_{-1/2}^{+1/2}\mbox{d}w\cdot [f(w)]^{\alpha_i}&=&
\int_{-1/2}^{+1/2}\mbox{d}w\cdot
[f(w)]^{\alpha_i}\cdot\left(\frac{|w|^\rho+D}{|w|^\rho+D}\right)^{\alpha_i}\\
&=&\int_{-1/2}^{+1/2}\mbox{d}w\cdot
[f(w)(|w|^\rho+D)]^{\alpha_i}\cdot\left[\frac{1}{(|w|^\rho+D)^{\theta_i}}\right]^{1-\alpha_i}\\
&\le&\left[\int_{-1/2}^{+1/2}\mbox{d}w\cdot
f(w)(|w|^\rho+D)\right]^{\alpha_i}\cdot
\left[\int_{-1/2}^{+1/2}\frac{\mbox{d}w}{(|w|^\rho+D)^{\theta_i}}\right]^{1-\alpha_i}\\
&=& (2D)^{\alpha_i}\cdot
\left[\int_{-1/2}^{+1/2}\frac{\mbox{d}w}{(|w|^\rho+D)^{\theta_i}}\right]^{1-\alpha_i}.
\end{eqnarray}
where $\theta_i=\alpha_i/(1-\alpha_i)$ and the third line follows from
H\"older's inequality. It remains to evaluate the integral
\begin{equation}
I=\int_{-1/2}^{+1/2}\frac{\mbox{d}w}{(|w|^\rho+D)^{\theta_i}}.
\end{equation}
To this end, we have to distinguish between the cases $\theta_i > 1/\rho$ and
$\theta_i < 1/\rho$ (the case $\theta_i=1/\rho$ can be solved separately or approached
as a limit of $\theta_i\to 1/\rho$ from either side). For the case $\theta_i >
1/\rho$, letting
\begin{equation}
c_i=\int_{-\infty}^{+\infty}\frac{\mbox{d}t}{(|t|^\rho+1)^{\theta_i}},
\end{equation}
we can easily bound $I$ as follows:
\begin{eqnarray}
I&=&D^{-\theta_i}\int_{-1/2}^{+1/2}\frac{\mbox{d}w}{(|w/D^{1/\rho}|^\rho+1)^{\theta_i}}\\
&\le&D^{1/\rho-\theta_i}\int_{-\infty}^{+\infty}
\frac{\mbox{d}(w/D^{1/\rho})}{(|w/D^{1/\rho}|^\rho+1)^{\theta_i}}\\
&\le&c_iD^{1/\rho-\theta_i}.
\end{eqnarray}
For $\theta_i < 1/\rho$, we proceed as follows:
\begin{eqnarray}
I&=&D^{1/\rho-\theta_i}\int_{-1/(2D^{1/\rho})}^{+1/(2D^{1/\rho})}
\frac{\mbox{d}t}{(|t|^\rho+1)^{\theta_i}}\\
&=&2D^{1/\rho-\theta_i}\int_0^{+1/(2D^{1/\rho})}\frac{\mbox{d}t}{(t^\rho+1)^{\theta_i}}\\
&\le&2D^{1/\rho-\theta_i}\int_0^{+1/(2D^{1/\rho})}\frac{\mbox{d}t}{(\max\{t^\rho,1\})^{\theta_i}}\\
&=&2D^{1/\rho-\theta_i}\int_0^{+1/(2D^{1/\rho})}\frac{\mbox{d}t}{\max\{t^{\rho\theta_i},1\}}\\
&=&2D^{1/\rho-\theta_i}\left[\int_0^{1}\frac{\mbox{d}t}{1}+
\int_1^{+1/(2D^{1/\rho})}\frac{\mbox{d}t}{t^{\rho\theta_i}}\right]\\
&=&2D^{1/\rho-\theta_i}\left[1+\frac{t^{1-\rho\theta_i}}{1-\rho\theta_i}\bigg|_1^{1/(2D^{1/\rho})}\right]\\
&=&2D^{1/\rho-\theta_i}\left[1+\frac{2^{\rho\theta_i-1}D^{\theta_i-1/\rho}-1}{1-\rho\theta_i}\right]\\
&\le&\frac{2^{\rho\theta_i}}{1-\rho\theta_i}.
\end{eqnarray}
Thus, defining $c_i'=2^{\alpha_i}\max\{c_i,2^{\rho\theta_i}/(1-\rho\theta_i)\}$, we
have
\begin{eqnarray}
\int_{-1/2}^{+1/2}\mbox{d}w\cdot [f(w)]^{\alpha_i}&\le&
(2D)^{\alpha_i}I^{1-\alpha_i}\\
&\le& c_i'\cdot D^{\zeta_\rho(\alpha_i)},
\end{eqnarray}
where the function $\zeta_\rho(\cdot)$ is defined as in (\ref{zeta}).
Thus,
\begin{equation}
U(f)\le c\cdot D^{\sum_{i=1}^k\zeta_\rho(\alpha_i)}
\end{equation}
where $c=\prod_{i=1}^kc_i'$.
Finally, it follows that
\begin{equation}
\tilde{R}(D)\ge -c\cdot D^{\sum_{i=1}^k\zeta_\rho(\alpha_i)}
\end{equation}
as claimed.


\end{document}